\documentclass[12pt,preprint]{aastex}

\usepackage{amsmath}
\usepackage{amssymb}
\usepackage{epsfig}

\newcommand{\kms}{km~s$^{-1}$}

\def\figurenum#1{\def\thefigure{#1}\let\@currentlabel\thefigure
    \addtocounter{figure}{\count22}}

\slugcomment{Last revised \today}

\shorttitle{Fit Quasar Spectrum CMA-ES}
\shortauthors{Wu \& Vanden Berk}

\begin{document}


\title{Fitting the Continuum Component of \\
    A Composite SDSS Quasar Spectrum Using CMA-ES}


\author{Jian Wu\altaffilmark{1,3}}
\email{fanchyna@gmail.com}

\and

\author{Daniel E. Vanden Berk\altaffilmark{2}}

\altaffiltext{1}{College of IST, Pennsylvania State University, University Park, PA, 16802}
\altaffiltext{2}{Department of Physics, Saint Vincent College, 300 Fraser Purchase Rd, Latrobe, PA, 15650}
\altaffiltext{3}{Department of Astronomy and Astrophysics, Pennsylvania State University, PA, 16802}


\begin{abstract}
Fitting the continuum component of a quasar spectrum in UV/optical band is 
challenging due to   
contamination of numerous emission lines. Traditional fitting
algorithms such as the least-square fitting and the Levenberg-Marquardt 
algorithm (LMA)
are fast but are sensitive to initial values of fitting parameters. 
They cannot guarantee to find global optimum solutions when the object functions
have multiple minima. In this work, we attempt to fit 
a typical quasar spectrum using the Covariance Matrix Adaptation Evolution 
Strategy (CMA-ES). The spectrum is
generated by composing a number of real quasar spectra
from the Sloan Digital Sky Survey (SDSS) quasar catalog data release 3 (DR3)
so it has a higher signal-to-noise ratio. 
The CMA-ES algorithm is an evolutionary algorithm that is designed to find 
the global rather than the local minima. The algorithm we implemented
achieves an improved fitting result than the
LMA and unlike the LMA, it is independent of initial parameter values. We are
looking forward to implementing this algorithm to real quasar 
spectra in UV/optical band. 
\end{abstract}




\section{Introduction}
We have constructed a set of 161 composite quasar
spectra binned in redshift and luminosity space
from $\sim80,000$ typical SDSS quasars.
Spectra in the same bin are normalized and averaged, so each composite
spectrum represents the \emph{average} properties of the spectra in each bin.
This is the first time this technique has been used to study the
evolution over a wide range of
luminosity ($38.25\lesssim\log{l_{\lambda}(2200\mbox{~\AA})}\lesssim44.00$)
and redshift ($0\lesssim z\lesssim5$).

We have tried two types of algorithms to derive a set of measurements from the
composite spectra.

\begin{figure}[t]
\centering
\includegraphics[angle=90,width=12cm]{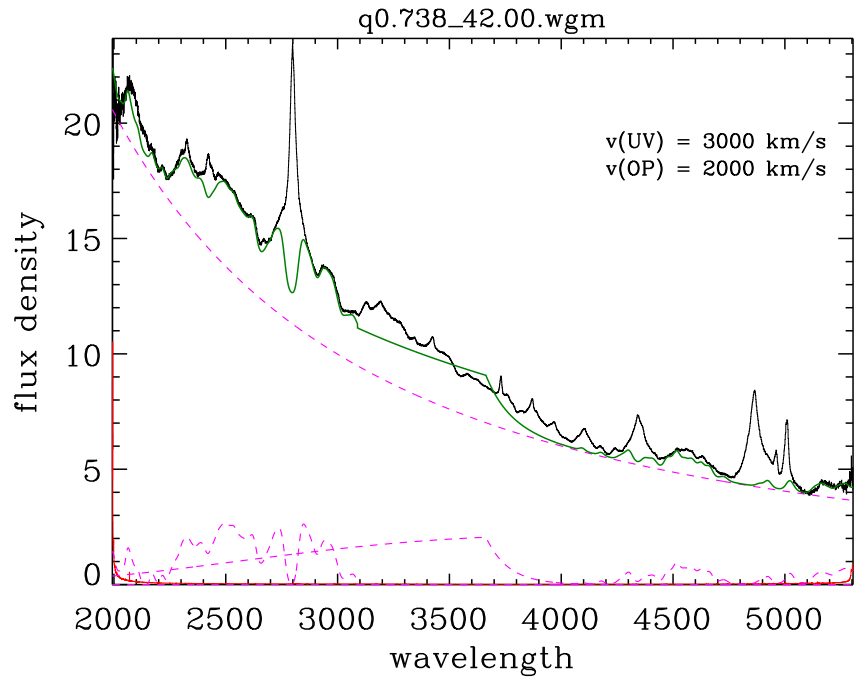}
\caption[Quasar spectrum fit by LMA.]{\mdseries The LMA fit to the
         SDSS composite quasar spectrum at $z=0.738$ and
         $\log{l_\lambda(\mbox{2200~\AA})}=42.00$. The original spectrum
         is shown in black. The $x$-axis is rest-frame wavelength in
         angstroms (\AA) and the $y$-axis is flux density in
            $10^{-17}$ erg~s$^{-1}$~cm$^{-2}$~\AA$^{-1}$. The green curve
            is the underlying continuum fit using the LMA; the dashed
            magenta lines are the individual continuum components. \label{fig1}}
\end{figure}

\begin{itemize}
  \item The Levenberg-Marquardt algorithm (LMA) is designed to solve the
        multivariate least-squares curve fitting problem \citep{jorge:guide}.
        It interpolates between the
        Gauss-Newton algorithm (GNA) \citep{bjorck:numerical} and
        the method of gradient descent \citep{avriel:2003},
        but it is more robust than the GNA, which means that in many cases, LMA
        finds a solution even if it starts far from the final minimum.
        The fitting results using this algorithm are displayed in
        Fig.~\ref{fig1}. This method has some limitations
        \begin{itemize}
          \item We found that to solve a problem with more than 10 free
                parameters, this algorithm is sensitive to initial values
                of parameters. We must provided an initial
                guess of the power-law (PL) component, and this estimate must
                be as close to the optimal value as possible.
          \item This algorithm does not work well for the iron templates,
                which do not have analytical expressions. The 
                iron emission and the small blue bump (SBB) cannot be
                fit well simultaneously (see the over-production of
                model flux around
                3700~\AA). In addition, we could not use this method to
                find the proper velocity dispersions of the iron template
                because the available dispersion values are discrete.
        \end{itemize}
  \item Genetic algorithm (GA). GA is a search heuristic that mimics the
        process of natural evolution \citep{ban98}.
        This heuristic is routinely used to
        generate useful solutions to optimization and search problems. A typical
        GA algorithm consists of initialization, selection, reproduction and
        termination. The solutions asymptotically converge to an optimal
        value under a given criteria.
        Vanden Berk wrote a program (not published) using this algorithm
        to fit the SDSS quasar spectra using GA, but this program has
        two disadvantages
        \begin{itemize}
          \item Instead of using the iron emission templates, each low
                ionization iron emission line is simulated using one
                or multiple Gaussian
                profiles. This requires over a hundred free parameters, which
                significantly slows down the fitting process. GA requires
                $\sim$5 -- 10 minutes (using a typical desktop computer
                with a 2.0~GHz processor) to fit a composite
                quasar spectrum.
          \item Although the program can produce a nearly perfect fit, the result
                is unstable. There are a number of parameter combinations that
                can produce equally good fits and it is difficult to
                determine which is the physical solution.
        \end{itemize}
\end{itemize}

The limitations and disadvantages of these two methods motivate us to develop an
evolutionary algorithm to fit the quasar spectrum. The CMA-ES is an
evolutionary algorithm for difficult non-linear non-convex optimization
problems in a continuous domain \citep{hansen:2004}. It is a second
order approach to estimate a positive definite matrix within an iterative
procedure (the covariance matrix). This approach makes the method feasible
on non-separable and/or ill conditioned problems. Because this method
does not require gradients, it is feasible on non-smooth and even
non-continuous problems, as well as ``noisy" problems. Previous results
using LMA imply that the space of the objective function is not smooth
as the solution varies depending on the initial guess \citep{hansen:2001}.
The CMA-ES approach can overcome this problem.

The paper is outlined in the following way. In
Section~\ref{problemdescription}, we describe the problem
in detail; in Section~\ref{solutionstrategy}, we present the solution strategy
in terms of each step applied in CMA-ES; in Section~\ref{programdesign}, we
briefly describe some important issues that arose when developing the
program; in Section~\ref{optimizationresults}, we present our
optimization results; and
finally in Section~\ref{discussionandfutureplan}, we discuss the limitations
of our approach and potential extension of this work.

\section{Problem Description}
\label{problemdescription}
We are aiming at fitting the underlying continuum of a composite quasar
spectrum at $z=0.738$, and $\log{l_\nu(2500\mbox{~\AA})}$ (see
Fig.~\ref{fig1}). This continuum is fit by four components
\begin{enumerate}
  \item A power-law (PL) continuum, which takes the form
        $f_\lambda=10^\beta\lambda^\alpha$.
        This component has two free parameters: the spectral
        index $\alpha$ and the scaling factor $\beta$. In principle, the
        variables do
        not have any constraints but previous studies have found that
        $\alpha\sim-1.6$ and $\beta\sim$6--7.
  \item A small blue bump (SBB), which takes the form
        \begin{equation}
          \label{eq-bbb}
          f_{\lambda}=\left\{\begin{array}{l}
            A_{\rm B}\cdot\frac{1-e^{\tau_{\rm B}}}{e^{hc/(k\lambda_{\rm B}T_{\rm B})}-1}\cdot e^{-(\lambda-\lambda_{\rm B})/\Delta\lambda},\lambda\geq\lambda_{\rm B}\\
            A_{\rm B}\cdot\frac{1-e^\tau}{e^{hc/(k\lambda T_{\rm B})}-1}\cdot \left(\frac{\lambda}{\lambda_{\rm B}}\right)^{-5},\lambda\leq\lambda_{\rm B},\tau=\tau_{\rm B}\left(\frac{\lambda}{\lambda_{\rm B}}\right)^3
          \end{array}\right.
        \end{equation}
        This component has three free parameters: Balmer temperature
        $T_{\rm B}$, Balmer optical depth $\tau_{\rm B}$, and scale factor
        $A_{\rm B}$. The scale factor should be positive; the other two
        parameters do not have constraints. Empirically, $\tau_{\rm B}\sim1$,
        and $T_{\rm B}\sim10^4$. Both $\lambda$ and $\lambda_{rm B}$ are in units
        of \AA\ and $\Delta\lambda=100$~\AA.
  \item UV iron emission forest. This is a blend of many low ionization iron
        emission lines ranging from $\sim1075$~\AA\ to 3090~\AA. This component
        has three free parameters, the scale factor $A_{\rm U}$, velocity
        dispersion $v_{\rm U}$ and relative shift $\Delta z_{\rm U}$\footnote{This is the shift of the template \emph{with respect to} the quasar spectrum itself, not the redshift of the quasar relative to the observer.}. The scale factor
        should be positive. The velocity dispersion can only be one of
        $1500,2000,\cdots9500$~km~s$^{-1}$ (set by a template grid).
        In principle, $\Delta z_{\rm U}$
        can be any value but typically $|\Delta z_{\rm U}|\leq0.005$.
  \item Optical iron emission forest. This is a blend of many low ionization
        iron emission lines ranging from $\sim3535$~\AA\ to $\sim7534$~\AA.
        Similar to the UV template, this component also has three free
        parameters, the scale factor $A_{\rm O}$, velocity dispersion
        $v_{\rm O}$ and relative shift $\Delta z_{\rm O}$. Again, $v_{\rm O}$
        can only be one of $1500,2000,\cdots,9500$~km~s$^{-1}$, and
        $|\Delta z_{\rm O}|\leq0.005$.
\end{enumerate}

A summary of the free parameters and their permitted values are listed in
Table~\ref{tab1}.
We use the \emph{reduced} $\chi^2$ value as the objective function:
$$\chi^2=\frac{1}{N-n-1}\sum_{i=1}^{N}\frac{\left(f_i-f_{e,i}\right)^2}{\sigma_i^2}$$
in which $N$ is the number of benchmark wavelength points, which
is used to compare the model and the observed fluxes; $n$ is the number of
variables; $f_i$ is the interpolated observed
flux value at wavelength point $i$; $f_{e,i}$ is the expected (calculated)
flux value at wavelength point $i$, and $\sigma_i$ is the interpolated
observational uncertainty at wavelength point $i$.
In this problem, we have $N=50$ benchmark points and $n=11$ variables.

\section{Solution Strategy}
\label{solutionstrategy}
We apply the CMA-ES as the method to solve this problem. This algorithm
involves sampling, selection, cumulation and updating operations
\citep{hansen:1996}.
The problem is initialized as Table~\ref{tab2} and each parameter shown
below such as $\mathbf{x}$, $\mathbf{m}$ and $\mathbf{z}$ are also
described in Table~\ref{tab2}. Each generation loop of
consists the following steps.

\begin{enumerate}
    \item Generate and evaluate $\lambda$ offspring
$$\mathbf{x}_i=\mathbf{m}+\sigma\mathbf{z}_i,\mathbf{z}_i\sim{\mathcal N}_i\left(\mathbf{0},\mathbf{C}\right),\quad i=1,2,\cdots,\lambda$$
  $$f_i=\chi^2\left(\mathbf{x}_i\right)$$
 Vector $\mathbf{m}$ has $N$ elements, and $\mathbf{C}$ is an $N\times N$ matrix
 (the covariant matrix). In this step, we apply a \emph{death penalty}
 to the two velocity dispersion values $v_{\rm U}$ and $v_{\rm O}$. If they
 fail to fall into the range of $[1500,9500]$, we simply discard this value
 and re-draw a new set of random numbers. However, if the number of
 failures is greater than 10, we stop drawing and adopt the boundary value.
 When evaluating the fitness function, we apply a \emph{penalty term}
 $f_{\rm p}(\mathbf{x})$ to
 the fitness function based on the value of redshift $\Delta z_{\rm U}$ and
 $\Delta z_{\rm O}$, so that
 $$\tilde{f}\left({\mathbf x}_i\right)=f\left({\mathbf{x}_i}\right)+f_{\rm p}\left({\mathbf x}_i\right)$$
 $$f_{\rm p}\left({\mathbf x}_i\right)=\left(C\cdot t\right)^2\cdot G\left({\mathbf x}_i\right)$$
 $$G\left({\mathbf x}_i\right)=\sum_{k=1}^4\max{\left[0,5\times g_k\left({\mathbf x}_i\right)\right]^2}$$
  The $g_k(\mathbf{x}_i)$ are the constrained items $[g_k({\mathbf x}_i)\leq0]$:
  \begin{equation*}
    \begin{array}{r@{=}l}
       g_1&\Delta z_{\rm U}-0.005\\
       g_2&-\Delta z_{\rm U}-0.005\\
       g_3&\Delta z_{\rm O}-0.005\\
       g_4&-\Delta z_{\rm O}-0.005
    \end{array}
  \end{equation*}
  In the equations above, $C=0.5$ and, $t$ is the generation number.

  \item Sort offspring by the penalized function $\tilde{f}({\mathbf x}_i)$ and
        compute the weighted mean. Although we sort the offspring using the
        penalized function, we still output the unpenalized function
        $f_{\rm p}({\mathbf x}_i)$, because it reflects the \emph{real}
        goodness-of-fit regardless of penalty term. In addition,
        $\tilde{f}({\mathbf x}_i)$
        involves the generation number $t$, so even if the best
        $\tilde{f}({\mathbf x}_i)$ decreases in the first few generations, it
        will increase after a certain point. After sorting the offspring
        ascendingly, we select the first $\mu$ offspring as the parents
        to calculate the weighted mean.
        The weighted mean after selecting $\mu$ offspring is
        $$m^{\rm s}_j=\sum_{i=1}^{\mu}w_ix^{\rm s}_{ij},\quad j=1,2,\cdots,N$$
        Note that $\mathbf{x}^{\rm s}$ is a matrix of $\mu\times N$ and
        $\mathbf{x}^{\rm s}$ is selected from $\mathbf{x}$ based
        on the fitness function so $x_{0j},j=1,2,\cdots,N$ contains the best
        offspring and thus has the heaviest weight $w_0$ (see appendix).
  \item  Cumulation: update the evolution paths.
         Conceptually, the evolution path is the path the strategy takes over
         a number of generation steps, which can be expressed as a sum of
         consecutive steps of the (weighted) mean $\mathbf{m}$.
         To accomplish this task, we iterate two path vectors,
         $\mathbf{p}_\sigma$ and
         $\mathbf{p}_c$, which represent the path of $\sigma$ and the
         covariance matrix $\mathbf{C}$. Both $\mathbf{p}_\sigma$ and
         $\mathbf{p}_c$ have two terms. The first term is the vector itself in
         the last iteration multiplied by a decay factor. This term causes
         the vector norm to decrease. The second term
         is based on the weighted mean of the selected offspring
         with a normalization factor and directs the evolution path
         to the optimal value. This step is a preparation to the covariance
         matrix adaption and step size adaption in the succeeding steps.
         The equations used to calculate these paths and the
          updates of evolution paths $\mathbf{p}_\sigma$ and $\mathbf{p}_c$,
          are shown below.
$$p^\sigma_j=(1-c_s)p^\sigma_j+\sqrt{c_s\left(2-c_s\right)\mu_{\rm eff}}\sum_{k=1}^NC^{-1/2}_{jk}z_k$$
  $$z_k=\frac{m_k^{\rm s}-m_k}{\sigma},\quad j=1,2,\cdots,N$$
 $$p^c_j=(1-c_c)p^c_j+h^\sigma\sqrt{c_c(2-c_c)\mu_{\rm eff}}z_j$$
 $$z_j=\frac{m_j^{\rm s}-m_j}{\sigma},\quad j=1,2,\cdots,N$$
 $$h^\sigma=\left\{\begin{array}{r@{,\quad}l}
           1 & \frac{||\mathbf{p}_c||}{N\sqrt{1-(1-c_s)^2t/\lambda}}<1.4+\frac{2}{1+N} \\
           0 & \frac{||\mathbf{p}_c||}{N\sqrt{1-(1-c_s)^2t/\lambda}}\geq1.4+\frac{2}{1+N}
 \end{array}\right.$$
  in which $C^{-1/2}_{jk}$ represents the array element in the inverse square
  root of matrix $\mathbf{C}$.
         \item  Adapt covariance matrix $\mathbf{C}$. This is the essential
          part of the
         algorithm and the reason why it is named as the covariance-matrix
         adaption evolutionary algorithm. The adaption equation includes
         three terms. The first term is the covariance matrix itself multiplied
         by a decay factor. Again, this term decreases the
         norm of the covariance matrix. The second term is the
         \emph{rank one update} plus
         a minor correction based on the evolutionary path $\mathbf{p}_c$;
         This term can learn straight ridges in $O(n)$ rather than $O(n^2)$
         function evaluations.
         The third term is the \emph{rank $\mu$ update} which is based on the
         weighted cross product of the selected offspring.
         The rank $\mu$ update increases the possible learning rate in
         large populations and also reduces the number of necessary
         generations roughly from
         $O(n^2)$ to $O(n)$ \citep{hansen:2003}. Therefore,
         the rank $\mu$ update is the primary
         mechanism whenever a large population size is used (say $\lambda>3n+10$).
         The equation for adaption of covariance matrix $\mathbf{C}$ is
$$C_{jk}=(1-c_{\rm l}-c_\mu)C_{jk}+c_{\rm l}\left[p^c_jp^c_k+(1-h^\sigma)c_c(2-c_c)C_{ij}\right]+\sum_{k=0}^Nc_\mu T_{jp}W_{pp}T_{pk}$$
in which ${\mathbf W}=\mbox{diag}\left(w_1,w_2,\cdots,w_\mu\right)$, $T_{jq}=x^{\rm s}_{jq}-M_{jq}, q=1,2,\cdots,\mu$, and $\mathbf{M}$ is an array of which
each column is $\mathbf{m}$. Note that $\mathbf{x}^{\rm s}$
represents the \emph{selected} and \emph{sorted} offspring.

  \item Finally, we update the step size $\sigma$. The reason
        for a step control is that the covariance matrix update
        can hardly increase the variance
        in \emph{all} directions simultaneously. In other words, the
        \emph{overall}
        scale of search, the \emph{global step-size} cannot be increased
        effectively \citep{ostermeier:1995}. In our investigation, we apply
        the \emph{path length control} (cumulative step-size adaption)
        as recommended by \cite{beyer:2003}. The equation for
  adaption of step size $\sigma$ is
$$\sigma=\sigma\exp{\left[\frac{c_s}{d_\sigma}\left(\frac{||\mathbf{p}_s||}{||\mathcal{N}(\mathbf{0},\mathbf{I})||}\right)-1\right]}.$$
  \item Correct $\mathbf{C}$. To ensure that $\mathbf{C}$ is
        a symmetric matrix, we enforce symmetry by replacing the lower
        triangle elements with the upper triangle elements.
  \item We apply four criteria to terminate the generation loop
      \begin{enumerate}
         \item The best fitness is below the pre-defined value.
         \item The number of generations is greater than the maximum number of
               generations allowed ($10^4$).
         \item The change of the best fitness value is smaller than 0.05 for
               20 consecutive generations.
         \item The change of the best fitness value is positive for 20
               consecutive generations. Theoretically, this could never
               happen, but in practice it does when the objective function approaches
               the optimal value.
      \end{enumerate}
\end{enumerate}
\section{Program Design}
\label{programdesign}
The coding design basically follows the flow of the loop in
Section~\ref{solutionstrategy}; here we briefly mention some
special aspects that arose when developing the code.
\begin{itemize}
  \item To avoid reading the iron emission templates from hard drive
        each time we
        evaluate the fitness function, we upload template spectra
        of \emph{all} the velocity dispersions
        before the generation loop. This requires a few seconds of
        some overhead computational time, but it significantly reduces the
        computational time required to evaluate the objective function.
        In each generation, we must evaluate
        the objective function at
        least 220 times (in the default situation).
  \item The available values of $v_{\rm U}$ and $v_{\rm O}$ are not
        continuous. As a result, we must perform a discretization
        to make sure that they are integer multiples of 500~\kms\ before
        evaluating the fitness function.
  \item To compute $\mathbf{C}^{-1/2}$, we first must
        find the eigenvectors of $\mathbf{C}$, and construct an array $\mathbf{P}$
        whose column vectors are these eigenvectors. Given that $\mathbf{C}$ is
        \emph{symmetric and positive deterministic},  it can be diagonalized as
        $\mathbf{D}=\mathbf{P}^{-1}\mathbf{C}\mathbf{P}$ whose diagonal elements
        are eigenvalues of $\mathbf{C}$. Therefore,
        $\mathbf{C}^{-1/2}=\mathbf{P}\left(\mathbf{D}^{1/2}\right)^{-1}\mathbf{P}^{-1}$.
  \item It is recommended to use the eigenvalues to construct the diagonal
        matrix ${\mathbf D}$, instead of obtaining ${\mathbf D}$ using matrix
        multiplication, \emph{i.e.},
        $\mathbf{D}={\mathbf P}^{-1}{\mathbf C}{\mathbf P}$. This is simply
        because of the limited precision of computers so that after a series of
        computations, the number zero is expressed in terms of a small number
        such as {\tt 1.1E-11}, but after a number of iterations, this number may
        accumulate, increase and eventually destroy the computation.
  \item Because of the termination criteria mentioned above, the last generation
        is not necessarily the best. As a result, after terminating the loop,
        the program reads the entire output file and selects the generation with
        the minimum fitness value as the final best result.
\end{itemize}
\section{Optimization Results}
\label{optimizationresults}
\subsection{Testing the Program}
\begin{figure}
  \centering
  \includegraphics[angle=90,width=12cm]{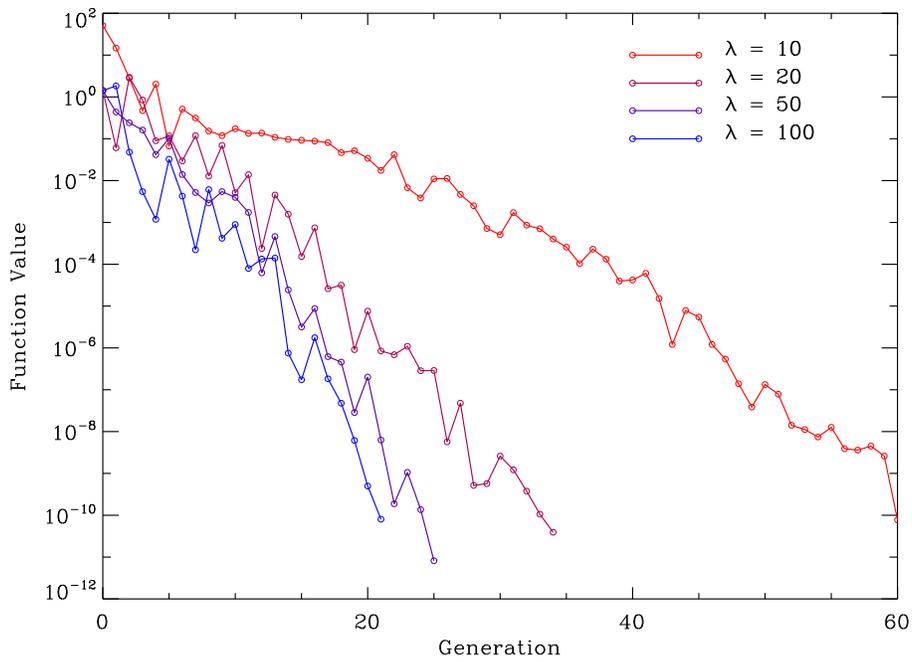}
  \caption[Rosenbrock function by varying population size]{
  \label{fig-testL}\mdseries Testing results of the Rosenbrock function
             by varying population size using CMA-ES. The colors of curves
            change from red to blue as the population size changes from 10
            to 100. We fix degrees of freedom (DOF) as 2. It is clear that
            a smaller population takes much more generations to converge
            than a large population. }
\end{figure}

Instead of using an existing code, we wrote our own code because
the objective function evaluation cannot be performed separately from the
optimization code. The code is written in interactive data language (IDL)
8.0 Mac version and run on a Mac Pro with two $2.66$~GHz Dual-Core Intel
Xeon processors and 2~GB 667 MHz memory.
Before using the code to fit the quasar spectrum,
we perform a series of tests by optimizing the generalized Rosenbrock
function \citep{rosenbrock:1960}:
$$f(\mathbf{x})=\sum_{i=1}^{n-1}\left[100\left(x_i^2-x_{i+1}\right)^2+(x_i-1)^2\right].$$
The global minimum of this function occurs when each $x_i$ equals
1 yielding $f(\mathbf{x})=0$. The global minimum is
inside a long, narrow parabolic-shaped flat valley.
This function is widely used as a test function because to
find the valley is trivial
but to converge to the global minimum is difficult \citep[][e.g.,]{storn:1997}.
The goal of this test is to prove that this code is working and achieves
the desired performance. We terminate the generation
loop when the function values decrease below $10^{-10}$.
We perform three tests. In the first test,
we fix the number of variables $n=2$ while changing the
population size $\lambda=10,20,50,100$. The evolution curves (Fig.~\ref{fig-testL})
indicate that the code is working and as the population size becomes
larger, the results converge faster.

\begin{figure}[h]
  \centering
  \includegraphics[angle=90,width=12cm]{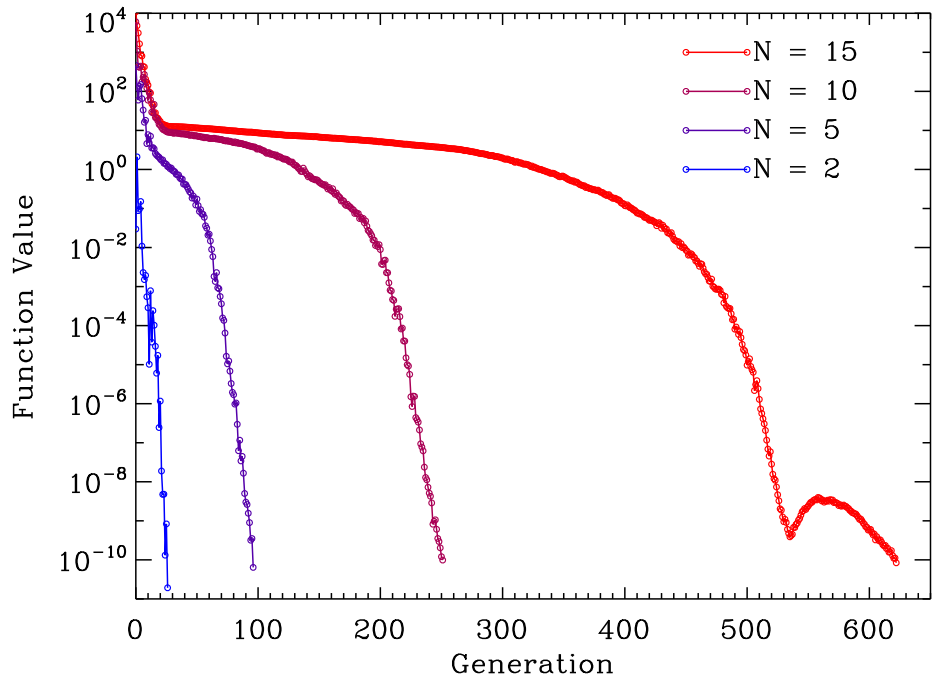}
  \caption[Rosenbrock function by varying DOF]{
  \label{fig-testN}\mdseries Testing results of the Rosenbrock function
            by varying DOF. The colors of the curves change from red to blue
as the DOF decreases from 15 to 2. We fix the population size as $\lambda=50$.
It is clear that an evolutionary process with a higher DOF converges much more
slowly than an evolutionary process with a lower DOF. }
\end{figure}

In the second test, we fix the population size at 50 and change the
number of variables $N=2,5,10,15$. The evolution curves
are presented in Fig.~\ref{fig-testN}. Again, these curves indicate
that the code is working well up to $N=15$, which is above the
size of the quasar spectrum fitting problem ($N=11$).
The number of generations requested to converge increases with problem size.

\begin{figure}[h]
  \centering
  \includegraphics[angle=90,width=12cm]{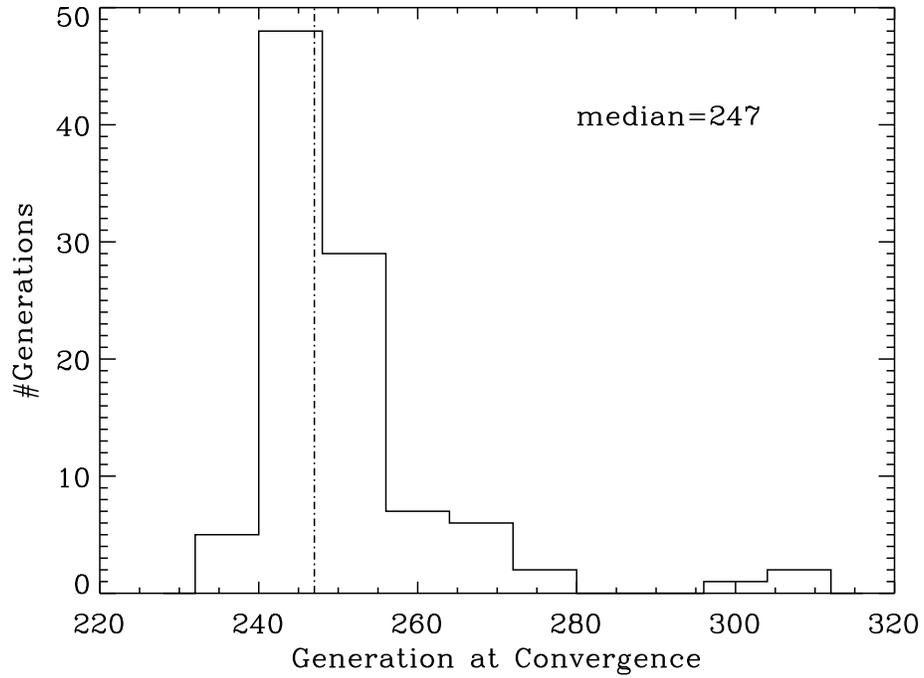}
  \caption[Rosenbrock function converge generation distribution]{
  \label{fig-testdist}\mdseries Distribution of the number of
     generations required for the Rosenbrock function to converge to $10^{-10}$
     using CMA-ES. We set the population size $\lambda=50$ and DOF
     $N=10$ and run the code 100 times. The median value (247) is presented
     as a dot-dashed line.}
\end{figure}

In the final test, we fix both population size $\lambda=50$ and DOF $N=10$,
and run the code 100 times. We want to determine (1) whether the
function can converge to the same value every run; (2) if it can, the
distribution of the number of generations it needs to converge
(Fig.~\ref{fig-testN}). We find that the objective function
is below $10^{-10}$ and $|x_i-1|<10^{-3}$ for all the runs.
The distribution resembles a Gaussian with a median of 247
(Fig.~\ref{fig-testdist}), and a dispersion of $\sim20$. Only
3 runs need more than $300$ generations.

The three tests verify that the code is working well for the
Rosenbrock function optimization problem up to $N=15$ with acceptable
reliability. In the following sections, we use this program to
fit the quasar spectrum.

\subsection{Results in Default Setup}

\begin{figure}[h]
  \centering
  \includegraphics[angle=90,width=12cm]{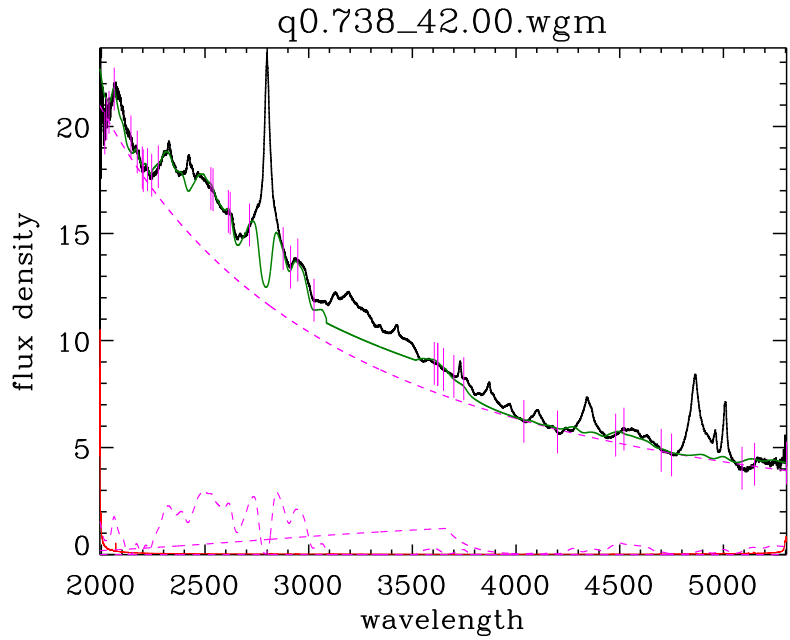}
  \caption[Spectral fitting using CMA-ES]{\label{fig2}\mdseries Fitting results using CMA-ES under the
           default setup. The legends are the same as Fig.~\ref{fig1},
           except that we over plot vertical lines in magenta at the
           benchmark wavelength points.}
\end{figure}

In the default setup, we fix the population size at $\lambda=220$ and the number
of selected offspring at $\mu=110$. The initial values of the free parameters are,
in general, of the same order of magnitude as the typical values (but
\emph{not} exactly the fitting results using other methods).
The optimization process terminates at the 72nd
generation. The values of basic parameters are presented in
Table~\ref{tab2}. The fitting results in the default setup are presented in
Fig.~\ref{fig2}. The optimal parameters we obtained are presented
and compared with previous results using LMA in Table~\ref{tab3}.
Comparing with the fitting result in Fig.~\ref{fig1}, we find a
significant improvement around the emission lines between 2500~\AA\ and
3000~\AA. Another evident improvement is the region around 3600~\AA\ and
3700~\AA. The LMA overproduces the flux around this wavelength range but
the CMA-ES produces a much better fit. However,
the CMA-ES over-produces the PL continuum between 4500~\AA\ and 5000~\AA.
In order to fit the broad bump (contributed by iron emission) around 4500~\AA,
the overall scale factor of the optical iron template is large so that the
total flux beyond 5000~\AA\ is over-produced. However, in general, the
CMA-ES algorithm produces a good balance between all the emission components
and the overall quality is indeed improved compared with the LMA method.
\subsection{Varying Parameters}
\begin{figure}[h]
  \centering
  \includegraphics[angle=90,width=10cm]{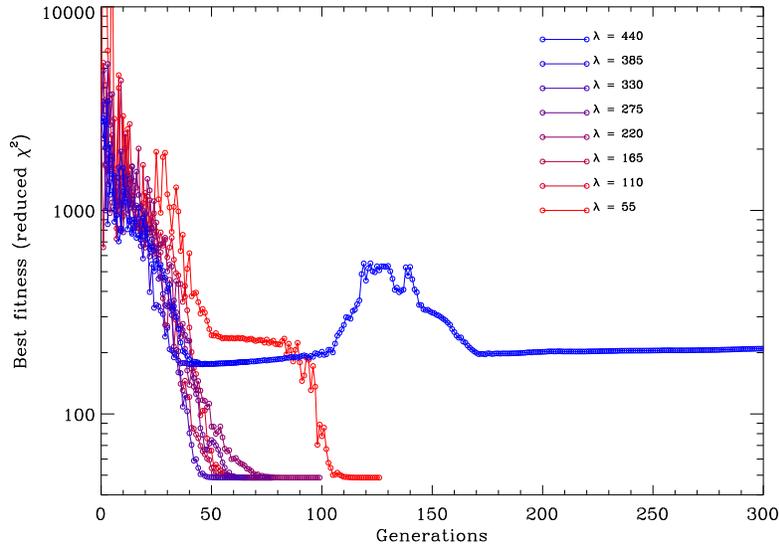}
  \caption[Evolutionary curves as the population size changes.]{\label{fig3}\mdseries
           Evolutionary curves as the population
           size changes. The colors of curves change from red to
           blue as the population increases from 55 to 440.
           The evolutionary curve at $\lambda=440$ does not coverage.
           The curve with $\lambda=55$ converges the most slowly;
           the curve with $\lambda=330$ converges the fastest.
 }
\end{figure}

In this section, we compare the performance of CMA-ES on this particular
problem by changing population size $\lambda$ and number of offspring $\mu$.
First, we fix the proportion of selected offspring with respect to the
entire population, which is 50\%, but change the population size.
The evolutionary curves are shown in Fig.~\ref{fig3} and the number
of generations for each population size is listed in Table~\ref{tab4}.

The population size changes
from 55 to 440 as the color changes from red to blue. We can see
that neither the red nor the blue curve has the best performance. The red
        curve, which represents the case $\lambda=55$, converges to the
        optimal values at $t\sim110$ but it is not the fastest option.
        The blue curve, which represents
        $\lambda=440$, actually does not converge to the optimal
        value; it bounces back at generation 100, returns at
        generation 170, and then remains around $f=110$. It even
        exhibits a gradual increase
        after $t\sim170$. The curve that shows the fastest convergence
        represents the case $\lambda=330$, which is 30 times the total number
        of free parameters.

\begin{figure}[h]
  \centering
  \includegraphics[angle=90,width=10cm]{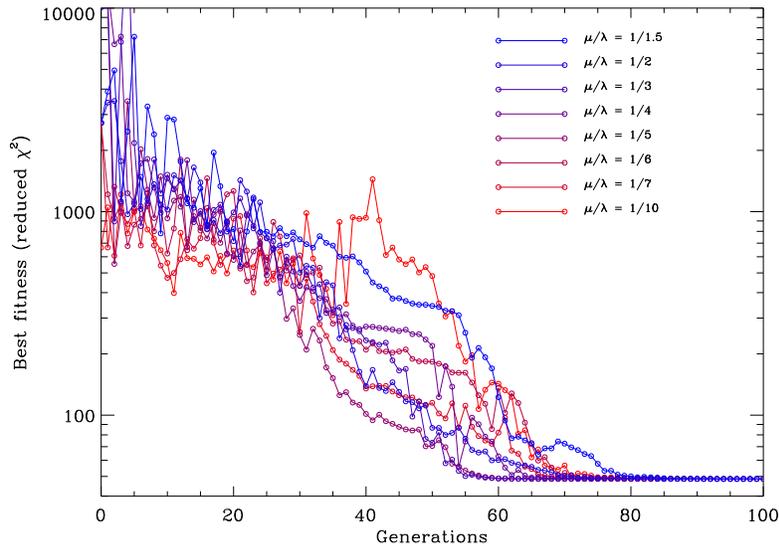}
  \caption[Evolutionary curves for different proportions of selected offspring with respect
           to population size]{\label{fig4}\mdseries Evolutionary curves for different
     proportions of selected offspring with respect to population size.
     The colors of curves change from red to blue as $\mu/\lambda$ increases
     from $1/10$ to $1/1.5$. These curves do not show a significant difference in
     convergence speed. }
\end{figure}

Next, we fix the population size $\lambda=22$, but change the proportion of
selected offspring. The evolutionary curves are
shown in Fig.~\ref{fig4} and the number of generation for each proportion
is listed in Table~\ref{tab5}. The color changes from red to blue as the
proportion increases from 1/10 to 1/1.5. From Fig.~\ref{fig4}, we do not see
a significant difference among these cases, although the case in which
$\mu/\lambda=1/3$ converges the fastest with the minimum number of
generations (64).

\begin{figure}[h]
  \centering
  \includegraphics[angle=90,width=10cm]{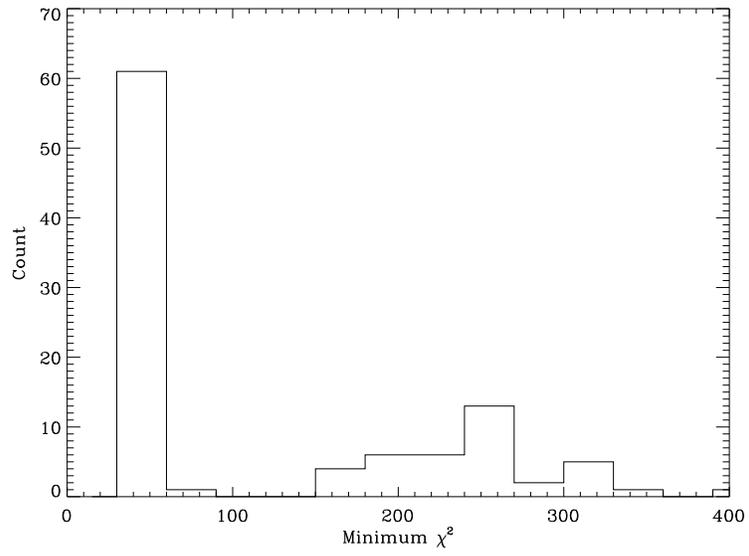}
  \caption[Distribution of $\chi^2$ values at termination]{\label{fig5}\mdseries
Distribution of $\chi^2$ values
         at termination in default setup $\lambda=220,\mu=110$.
         About 60\% of cases converge (the high bar on the left).}
\end{figure}

Finally, we test the reliability of CMA-ES on
the quasar spectrum fitting problem by running the program under the
default setups ($\lambda=220$, $\mu=110$) 100 times.
The distribution of the number of generations at which the evolution terminates
is represented in Fig.~\ref{fig5}. This
figure illustrates that in general the CMA-ES algorithm is reliable;
about 60\% of runs converge to the optimal value.

\section{Conclusion and Discussion}
\label{discussionandfutureplan}
The CMA-ES method has achieved a preliminary success in the continuum fitting
problem of a typical composite quasar spectrum. In the best case, it can finish
fitting a spectrum in about 1 minute under the default setups. Although this is
about 5 times longer than the LMA algorithm, it is still within the
acceptable time scale. The most important aspect is that \emph{the overall
fitting quality is significantly improved} and \emph{the optimal
results do not depend on the initial parameters}.
It is then feasible to extend its application to more quasar spectra and
more complicated cases.

The real case can be more complicated. For instance, an iron
emission template (either UV or optical) can be subdivided into a
number of sections and each section may have a different scale factor.
For some objects, a single PL is not enough and we usually need
to include another PL at $\lambda>5600$~\AA\ (if covered).
Because of effective exposure time, spectra may have
different signal-to-noise ratio. All of these factors may complicate
the fitting process. Although the task in this work is simplified,
the spectrum we are using for the test is representative of the entire
sample.

The fact that the theoretical reduced $\chi^2$ value ($\chi^2\sim1$)
is not reached
is not because of the algorithm but the model used the fit the
spectrum. A single power-law may not be an accurate description to the
overall spectral profile. The iron emission shape may vary with different
quasars, so a single iron emission template may not be exact. The emission
lines can also contribute some flux at the benchmark wavelength points
so it may not be correct to only count the contribution from the continuum
at these wavelength points. Another issue is the non-uniformity of the
benchmark wavelength points. When selecting these wavelength points, we
were trying to use \emph{as many} constraints as possible. However, because
the benchmark points must avoid \emph{emission lines} and there is
a region in which the iron emission template is \emph{not available}
(between 3000~\AA\ and 3500~\AA), these benchmark points are not uniformly
distributed; this could affect the evaluation.

Despite these limitations, the CMA-ES produces a limit we can
reach under the best fitting model
we can provide. It provides an acceptable fitting of a quasar spectrum
in a timely manner. Based on these results, we
can further fit the \emph{emission lines} seen in the spectra.

\acknowledgments
I thank Dr. Patrick Reed for his suggestions and comments on 
this project midterm review. 

\clearpage

\begin{deluxetable}{c|l|c|c|c}
\tablecolumns{5}
\tablewidth{0pc}
\tabletypesize{\footnotesize}
\tablecaption{\mdseries Spectral components and free parameters.\label{tab1}}
\tablehead{
   \colhead{N}                    &
   \colhead{Name}                 &
   \colhead{Expression}           &
   \colhead{Free Para.}           &
   \colhead{Constraint}
}
\startdata
              &    & & $\alpha$ & $\alpha\sim-1.6$\\
   {1} & Power-law (PL) & analytical & $\beta$  & $\beta\sim6$--$7$\\
\hline
    &  & & $A_{\rm B}$ & $A_{\rm B}>0$ \\
 2  & Small Blue Bump (SBB) & analytical\tablenotemark{1} & $T_{\rm B}$&  $T_{\rm B}\sim10^4$ \\
    &  &  & $\tau_{\rm B}$& $\tau_{\rm B}\sim1$ \\
\hline
    &           &              & $A_{\rm U}$ & $A_{\rm U}>0$\\
 3  & UV iron emission& templates & $v_{\rm U}$  &  $v_{\rm U}=1500,2000,\cdots,9500$ \\
    &     &     & $\Delta z_{\rm U}$ & $|\Delta z_{\rm U}|\leq0.005$ \\
\hline
  &            &             & $A_{\rm O}$ & $A_{\rm O}>0$\\
 4 &  Optical iron emission & templates & $v_{\rm O}$ &  $v_{\rm O}=1500,2000,\cdots,9500$ \\
   &   & & $\Delta z_{\rm O}$ &  $|\Delta z_{\rm O}|\leq0.005$ \\ \hline
\enddata
\tablenotetext{1}{The analytical form is seen in Equation~\ref{eq-bbb}.}
\end{deluxetable}

\begin{deluxetable}{c|l|l}
\tablecolumns{3}
\tablewidth{0pc}
\tabletypesize{\footnotesize}
\tablecaption{\mdseries Initialization. Values of $\mathbf{m}$ are typical values from previous results.\label{tab2}}
\rotate
\tablehead{
  \colhead{Para.}               &
  \colhead{Initialization}      &
  \colhead{Remarks}
}
\startdata
             &   & (weighted) means of parameters, including \\
$\mathbf{m}$ &  $[-1.6,7,150,10000,1,0.02,3000,0,0.005,3000,0]$ & $[\alpha,\beta,A_{\rm B},T_{\rm B},\tau_{\rm B},A_{\rm U},v_{\rm U},\Delta z_{\rm U},A_{\rm O},v_{\rm O},\Delta z_{\rm O}]$\\ \hline
$\lambda$     &  $20\times N=220$           &  population size \\ \hline
$\mu$     &  $\lambda/2=110$          &    number of parents for recombination \\ \hline
${\mathbf w}$     & $w_j=\log{(\mu+1/2)}-\log{j},\quad j=1,2\cdots,\mu$ & normalized weights \\ \hline
$\mu_{\rm eff}$     &   $1/\sum{w_i^2}=57.6986$    & variance-effectiveness    \\ \hline
$\mathbf{C}$ & $C_{jk}=1(j=k),=0(j\neq k)\quad j,k=1,2,\cdots,N$ & covariance matrix\\ \hline
$c_c$ & $(4+\mu_{\rm eff}/N)/(N+4+2\mu_{\rm eff}/N)=0.3627$  & time constraint for cumulation for $\mathbf{C}$ \\ \hline
 $c_\sigma$  & $(\mu_{\rm eff}+2)/(N+\mu_{\rm eff}+5)=0.81$ & const for cumulation for $\sigma$ control\\ \hline
$c_{\rm l}$ & $2/\left[(N+1.3)^2+\mu_{\rm eff}\right]=0.01$    & learning rate for rank-one update of $\mathbf{C}$\\ \hline
$c_{\mu}$     & $2(\mu_{\rm eff}-2+1/\mu_{\rm eff})/\left[(N+1)^2+\mu_{\rm eff}\right]=0.49$ & learning rate for rank-$\mu$ update  \\ \hline
 $d_\sigma$ & $1+2\max{\left(0,\sqrt{\left[(\mu_{\rm eff}-1)/(N-1)\right]-1}\right)}=4.157$  & damping for $\sigma$          \\ \hline
 $\mathbf{p}^{\rm c}$ & $p^c_j=0,\quad j=1,2,\cdots,N$   & evolution paths for $\mathbf{C}$   \\ \hline
 $\mathbf{p}^{\sigma}$ & $p^\sigma_j=0,\quad j-1,2,\cdots,N$   & evolution paths for $\sigma$   \\ \hline
 $||\mathcal{N}(\mathbf{0},\mathbf{I})||$ & $\sqrt{N}\left[1-1/(4N)+1/(21N^2)\right]=3.24$ & expectation of $||\mathcal{N}(\mathbf{0},\mathbf{I})||$\\ \hline
\enddata
\end{deluxetable}

\begin{table}[h]
  \centering
  \caption{\label{tab3}\mdseries Optimal parameter values using LMA and
          CMA-ES (under the default setup).}
          \vspace{5mm}
  \begin{tabular}{c|l|l}
   \hline\hline 
   Para. & LMA & CMA-ES\\ \hline
   $\alpha$ & -1.768 & -1.713\\ \hline
   $\beta$  & 7.147 & 6.974\\  \hline
   $A_{\rm B}$ & 134.827 & 150.681\\ \hline
   $\tau_{\rm B}$ & 1 & 0.5153\\ \hline
   $T_{\rm B}$ & 10000 & 10001\\ \hline
   $A_{\rm U}$ & 0.028 & 0.044\\ \hline
   $v_{\rm U}$ & 3000 & 3000\\ \hline
   $\Delta z_{\rm U}$ & 0.0000 & -0.0008\\ \hline
   $A_{\rm O}$ & 0.015  & 0.015\\ \hline
   $v_{\rm O}$ & 3000 & 3000\\ \hline
   $\Delta z_{\rm O}$ & 0.000 & -0.0046\\ \hline
  \end{tabular}
\end{table}

\begin{table}[h]
  \centering
  \caption{\label{tab4}\mdseries Termination generation as
          a function of population size $\lambda$. The proportion of
          selected offspring is fixed to be 50\%.}
        \vspace{5mm}
  \begin{tabular}{c|c}
   \hline \hline
    Population & Number of \\
    Size       & Generation \\ \hline
    55 & 127 \\
    110 & 100 \\
    165 & 82 \\
    220 & 72 \\
    275 & 78 \\
    330 & 66 \\
    385 & 76 \\
    440 & NA \\ \hline
  \end{tabular}
\end{table}

\begin{table}[h]
  \centering
  \caption{\label{tab5}\mdseries Number of generations as a function of
           the proportion of selected offspring with respect to the
          fixed total population size $\lambda=220$.}
          \vspace{5mm}
  \begin{tabular}{c|c}
   \hline\hline
                & Number of \\
     Proportion & Generations \\ \hline
     1/10 & 101 \\
     1/7 & 81 \\
     1/6 & 94 \\
     1/5 & 82 \\
     1/4 & 69 \\
     1/3 & 64 \\
     1/2 & 72 \\
     1/1.5 & 102 \\ \hline
  \end{tabular}
\end{table}

\bibliographystyle{apj}
\bibliography{ref}

\begin{thebibliography}{12}
\expandafter\ifx\csname natexlab\endcsname\relax\def\natexlab#1{#1}\fi

\bibitem[{Andreas~{Ostermeier}(1995)}]{ostermeier:1995}
Andreas~{Ostermeier}, A. G. . N.~H. 1995, Evolutionary Computation, 2, 369

\bibitem[{Avriel(1993)}]{avriel:2003}
Avriel, M. 1993, ACM Trans. Program. Lang. Syst., 15, 745

\bibitem[{Banzhaf {et~al.}(1998)Banzhaf, Francone, Keller, \& Nordin}]{ban98}
Banzhaf, W., Francone, F.~D., Keller, R.~E., \& Nordin, P. 1998, Genetic
  programming: an introduction: on the automatic evolution of computer programs
  and its applications (San Francisco, CA, USA: Morgan Kaufmann Publishers
  Inc.)

\bibitem[{Beyer \& Arnold(2006)}]{beyer:2003}
Beyer, H.-G., \& Arnold, D.~V. 2006, Evolutionary Computation, 11, 19

\bibitem[{Bj{\"o}rck(1996)}]{bjorck:numerical}
Bj{\"o}rck, {\AA}. 1996, Numerical Methods for Least Squares Problems
  (Philadelphia: SIAM)

\bibitem[{Hansen \& Ostermeirer(1996)}]{hansen:1996}
Hansen, \& Ostermeirer. 1996, in Proceedings of the 1996 IEEE International
  Conference on Evolutionary Computation, 312--317

\bibitem[{Hansen \& Koumoutsakos(2003)}]{hansen:2003}
Hansen, M., \& Koumoutsakos. 2003, Evolutionary Computation, 11, 1

\bibitem[{Hansen \& Kern(2004)}]{hansen:2004}
Hansen, N., \& Kern, S. 2004, in Eighth International Conference on Parallel
  Problem Solving from Nature PPSN VIII, Proceedings, Berlin: Springer,
  282--291

\bibitem[{Hansen \& Ostermeier(2001)}]{hansen:2001}
Hansen, N., \& Ostermeier, A. 2001, Evolutionary Computation, 9, 159

\bibitem[{More \& Wright(1993)}]{jorge:guide}
More, J.~J., \& Wright, S.~J. 1993, Optimization Software Guide (Philadelphia,
  PA, USA: Society for Industrial and Applied Mathematics)

\bibitem[{Rosenbrock(1960)}]{rosenbrock:1960}
Rosenbrock, H.~H. 1960, Computer J., 3, 175

\bibitem[{Storn \& Price(1997)}]{storn:1997}
Storn, R., \& Price, K. 1997, Journal of Global Minimization, 11, 341

\end{thebibliography}

\end{document}